\documentclass[twoside]{article}
\usepackage{bm,amsmath,amssymb}
\usepackage{indentfirst}
\usepackage{multicol}

\def\be{\begin{equation}}
\def\ee{\end{equation}}
\def\bea{\begin{eqnarray}}
\def\eea{\end{eqnarray}}
\def\nn{\nonumber}
\def\p{\partial}

\newcommand\no{\noindent}
\newcommand\bc{\begin{center}}
\newcommand\ec{\end{center}}
\newcommand\bmu{\begin{multicols}{2}}
\newcommand\emu{\end{multicols}}
\newcommand\vs{\vspace}

\newcommand\hs{\hspace}
\newcommand\fo{\footnotesize}
\newcommand\yj[1]{{$^*$}\parbox[t]{168mm}{#1}}
\newcommand\gj[1]{\parbox[t]{133mm}{\normalsize #1}}

\catcode`@=11\def\@evenhead{\vbox{\hbox to
\textwidth{\footnotesize\rm\thepage\qquad\qquad\ \ \hfill
{Peng Jun-Jin \sl et al} \hfill 
}\vs{1mm} \hbox
to\textwidth{\hfill\vrule depth0pt height0.15mm width\textwidth\hfill}}}
\def\@oddhead{\vbox{\hbox to \textwidth{\footnotesize\rm 
\hfill Hawking radiation from the Schwarzschild black hole ...
\hfill\thepage} \vs{1mm} \hbox to \textwidth{\vrule depth0pt
height0.15mm width\textwidth}}}

\begin{document}

\setcounter{page}{1}
\thispagestyle{empty}

\begin{flushright}
arXiv: 0705.1225v3 [hep-th]
\end{flushright}

\bc
{\LARGE\bf Hawking radiation from the Schwarzschild black hole with a global monopole
via gravitational anomaly$^{*}$} \footnotetext{\hs*{-.45cm}\fo\yj{Project supported
by the National Natural Science Foundation of China (Grant No 10675051).} \\
\hs*{.8mm}$^\dag$Corresponding author. E-mail: pengjjph@163.com \ \ $^\natural$ sqwu@phy.ccnu.edu.cn}

\vs{5mm}

{\rm Peng Jun-Jin$^{*}$ \ \ and Wu Shuang-Qing$^\natural$}

\vs*{2mm}

{\fo\sl College of Physical Science and Technology, Central China Normal University,
Wuhan, 430079}

\vs{3mm}
{\fo (Revised 11 July 2007)}

\vs{4mm}

\parbox[c]{155mm}{\parindent 20pt\fo
Hawking flux from the Schwarzschild black hole with a global monopole is obtained by using
Robinson and Wilczek's method. Adopting a dimension reduction technique, the effective
quantum field in the $(3+1)$--dimensional global monopole background can be described by an
infinite collection of the $(1+1)$--dimensional massless fields if neglecting the ingoing modes
near the horizon, where the gravitational anomaly can be cancelled by the $(1+1)$--dimensional
black body radiation at the Hawking temperature.

\vs{3mm}

\no{\normalsize \bf Keywords:}\ \gj{anomaly, Hawking radiation, black hole, global monopole}

\no{\normalsize \bf PACC:}\normalsize\ 04.\,70.\,Dy, 04.\,62.\,+v, 11.\,30.\,-j
}
\ec

\vs{3mm}
\bmu

Recently, Robinson and Wilczek$^{[1]}$ (RW) proposed an intriguing approach to derive Hawking
radiation from a Schwarzschild-type black hole through gravitational anomaly. Their basic idea
goes as follows. Consider a massless scalar field in the higher dimensional space--time. Upon
performing the dimensional reduction technique together with a partial wave decomposition, they
found that the physics near the horizon in the original black hole background can be described
by an infinite collection of the massless field in a $(1+1)$--dimensional effective field theory.
When omitting the classically irrelevant ingoing modes in the region near the horizon, the effective
theory becomes chiral and there exist gravitational anomalies in the near--horizon region, which
just can be cancelled by the $(1+1)$--dimensional black body radiation at the Hawking temperature.
As is shown later, the RW's method is very universal, and soon was extended to other black hole
cases$^{[2-5]}$ which contain gauge anomaly in addition to gravitational anomaly.

In this paper, we will use the RW's method to investigate Hawking radiation of a static spherically
symmetric black hole with a global monopole from the viewpoint of cancelling the gravitational anomaly.
During the process of the GUT phase transition, we imagine that a Schwarzschild black hole swallows a
global monopole, forming a black-hole--global-monopole system. An unusual and stirring property of this
black-hole--global-monopole system$^{[6,7]}$ is that it possesses a solid deficit angle, which makes it
quite different topologically from that of a Schwarzschild black hole alone. Thermodynamical properties
of such a static spherically symmetric system have been studied extensively in Ref.~[7]. Because the
background space--time considered here is not asymptotically flat rather it contains a topological defect
due to the presence of a global monopole, the RW's method, however, cannot be directly applied for the
black-hole--global-monopole system. In the following, we shall adopt a slightly different procedure and
perform various coordinate transformations before we can use the RW's method. Accordingly, we generalize
the RW's method to the more general case that the $g_{tt}$ and $g_{rr}$ components of the metric satisfy
$g_{tt}\cdot g_{rr}\neq 1$.

The metric of a Schwarzschild-type black hole with the global $O(3)$ monopole is described by$^{[6,7]}$
\be\begin{aligned}
ds^2 &= -f(r)dt^2 +f(r)^{-1}dr^2 +r^2d\Omega_2^2 \, , \\
f(r) &= 1 -\eta^2 -2m/r \, ,
\end{aligned}\ee
where $m$ is the mass parameter of the black hole and $\eta$ is related to the symmetry breaking
scale when the global monopole is formed during the early universe. 
For a typical GUT symmetry breaking scale, $\eta^2 \sim 10^{-6}$, so it's reasonable to assume
$1 -\eta^2 \simeq 1$ throughout this paper.

Since the prime physical quantity obtained by means of the RW's method is the Hawking temperature
which enters into the first law of black hole thermodynamics, let's begin with by reviewing the
thermodynamical properties of the black-hole--global-monopole system.

For the space--time metric (1), the Hawking temperature and the entropy are given by$^{[7]}$
\be\begin{aligned}
T &= \frac{\kappa}{2\pi} = \frac{\p_rf\big|_{r_H}}{4\pi\sqrt{1 -\eta^2}}
  = \frac{(1-\eta^2)^{3/2}}{8\pi m} \, , \\
S &= \frac{A}{4} = \pi r_H^2 = \frac{4\pi m^2}{(1 -\eta^2)^2} \, ,
\end{aligned}\ee
where $\kappa$ and $A$ are, respectively, the surface gravity and the area at the horizon
$r_H = 2m/(1 -\eta^2)$.

The Arnowitt-Deser-Misner (ADM) mass $M$ of the system can be calculated via the Komar integral
\be
M = \frac{-1}{8\pi}\oint{\xi_{(t)}^{\mu;\nu}d^2\Sigma_{\mu\nu}}
  = \frac{m}{\sqrt{1 -\eta^2}} \, ,
\ee
where $\xi_{(t)}^{\mu} = (1 -\eta^2)^{-1/2}(\p_t)^{\mu}$ is the normalized time--like Killing
vector. Obviously, the ADM mass $M$ isn't equal to the mass parameter $m$ because of the presence
of a global monopole. One can easily show that the ADM mass $M$, the temperature $T$ and the
entropy $S$ given above obey the differential and integral forms of the first law of black hole
thermodynamics as follows
\be
dM = TdS \, , \qquad M = 2TS \, .
\ee

Now introducing the following coordinate transformation
\be
t \to (1 -\eta^2)^{1/2}t \, , \quad r \to (1 -\eta^2)^{-1/2}r \, ,
\ee
and defining a new mass parameter
\be
\widetilde{m} = (1 -\eta^2)^{-3/2}m \, ,
\ee
then we can rewrite the line element (1) as
\be\begin{aligned}
ds^2 &= -f(r)dt^2 +f(r)^{-1}dr^2 +(1 -\eta^2)r^2d\Omega_2^2 \, , \\
f(r) &= 1 -2\widetilde{m}/r \, .
\end{aligned}\ee
This metric is, apart from the deficit solid angle $4\pi\eta^2$, very similar to the Schwarzschild
solution. Remarkably, due to the presence of a global monopole, the original space--time (1) is not
asymptotically flat but asymptotically bounded. After performing the above coordinate transformation,
which is a scale transformation, the metric is brought into an asymptotic one although containing a
topological defect. The advantage of this transformation is to make the calculation of the ADM mass
and the Hawking temperature more feasible.

For the line element (7), the surface gravity at the horizon $r_H = 2\widetilde{m}$ can be determined
as $\kappa = \frac{1}{2}\p_rf\big|_{r_H}$. Analogously, one can compute the ADM mass $M = (1 -\eta^2)
\widetilde{m}$, the Hawking temperature $T = \kappa/(2\pi) = 1/(8\pi\widetilde{m})$, and the entropy
$S = A/4 = 4\pi(1 -\eta^2)\widetilde{m}^2$, and find that they are essentially identical to those
given by Eqs. (2) and (3) by virtue of the relation (6), thus satisfying the same Bekenstein-Smarr's
relationship (4).

In the following, we will apply the RW's method to show that the flux of Hawking radiation from
a Schwarzschild-type black hole with the global monopole can be determined by anomaly cancellation
conditions and regularity requirement at the horizon. The RW's method, however, cannot be immediately
applied to obtain the correct formula of Hawking temperature for the line element (1), rather it can
be directly used to obtain that for the metric (7). Thus, we shall first base our analysis below
upon the metric (7) but will soon turn to the space--time (1).

For simplicity, let's consider the action for a massless scalar field in the background space--time
(7). After performing the partial wave decomposition $\varphi = \sum_{lm}\varphi_{lm}(t, r)Y_{lm}(\theta,
\phi)$, and only keeping the dominant terms, the action becomes
\bea
&&\hspace*{-0.4cm} S[\varphi]
 = -\frac{1}{2}\int d^4x\sqrt{-g} g^{\mu\nu}\p_{\mu}\varphi\p_{\nu}\varphi \nn \\
&&\quad = \frac{1}{2}\int dtdrd\theta d\phi (1 -\eta^2)r^2\sin\theta
 \varphi\Big[\frac{-1}{f}\p_t^2 \nn \\
&&\qquad +\frac{1}{r^2}\p_r\big(r^2f\p_r\big)
 +\frac{1}{(1 -\eta^2)r^2}\Delta_{\Omega}\Big]\varphi \nn \\
&&\quad \approx \frac{1}{2}\sum_{lm}\int dtdr (1 -\eta^2)r^2\varphi_{lm} \nn \\
&&\qquad~ \times \Big[\frac{-1}{f}\p_t^2 +\p_r\big(f\p_r\big)\Big]\varphi_{lm} \, ,
\eea
where $\Delta_{\Omega}$ is the angular Laplace operator. Apparently, a free scalar field in the
original $(3+1)$--dimensional background can be effectively described by an infinite collection
of massless fields in the $(1+1)$--dimensional space--time with the metric
\be ds^2 = -f(r)dt^2 +f(r)^{-1}dr^2 \, ,
\ee
together with the dilaton field $\Psi = (1 -\eta^2)r^2$. On the other hand, if we start with the
metric (1) and perform the dimension reduction, the same effective metric yields but with a different
dilaton factor $\Psi = r^2$ and $f(r)$ is now given by (1).

Yet we can still go beyond further. For later usage, consider now the most general, static and
spherically symmetric black hole solutions with the line element
\be
ds^2 = -f(r)dt^2 +h(r)^{-1}dr^2 +P(r)^2d\Omega_2^2 \, .
\ee
A similar dimension reduction technique leads to the $(1+1)$--dimensional effective metric
\be
ds^2 = -f(r)dt^2 +h(r)^{-1}dr^2 \, ,
\ee
with the dilaton field $\Psi = P(r)^2$, which makes no contribution to the anomaly. At the horizon
$r = r_H$, the surface gravity is $\kappa = \frac{1}{2}\sqrt{f^{\prime}(r_H)h^{\prime}(r_H)}$,
where a prime denotes the derivative with respect to $r$.

Since we are considering a static background, the contribution from the dilaton field can be
neglected. Thus we find that the physics near the horizon can be described by an infinite
collection of massless fields in the $(1+1)$--dimensional effective theory on the background
metric (11).

Next we turn to the gravitational anomaly. A gravitational anomaly is an anomaly in the general
coordinate covariance, taking the form of non-conservation of energy-momentum tensor. The consistent
one arising in the $(1+1)$--dimensional chiral theory reads
\be
\nabla_{\mu}T_{~\nu}^{\mu} = \frac{1}{96\pi\sqrt{-g}}
\varepsilon^{\beta\delta}\p_{\delta}\p_{\alpha}\Gamma_{\nu\beta}^{\alpha} \, ,
\ee
on the other hand, the covariant anomaly for outgoing modes reads
\be
\nabla_{\mu} \widetilde{T}_{~\nu}^{\mu} =
\frac{-1}{96\pi\sqrt{-g}}\varepsilon_{\mu\nu}\p^{\mu}R \, ,
\ee
where $\varepsilon^{\mu\nu}$ is an antisymmetric tensor with $\varepsilon^{tr} = 1$.

We will localize the physics outside the horizon since the effective theory is defined in the
exterior region $[r_H, +\infty]$. Now we divide the region outside the horizon into two parts:
the near--horizon region $[r_H, r_H +\varepsilon]$, where the effective quantum field theory
becomes chiral and exhibits a gravitational anomaly, and the other region $[r_H +\varepsilon,
+\infty]$, where the theory is not chiral and there is no anomaly. So let's focus on the anomaly
in the region $[r_H, r_H +\varepsilon]$. Having omitted the classically irrelevant ingoing modes
near the horizon, the energy--momentum tensor in this region exhibits an anomaly, which can be
written as
\be
\nabla_{\mu}T_{(H)\nu}^{\mu} \equiv A_{\nu} \equiv \frac{1}{\sqrt{-g}}\p_{\mu}N_{~\nu}^{\mu} \, .
\ee
For a metric of the form (11), $N_{~\nu}^{\mu} = A_{\nu} = 0$ in the region $[r_H +\varepsilon, +\infty]$.
But in the near--horizon region $[r_H, r_H +\varepsilon]$, the components of $N_{~\nu}^{\mu}$ are
\be\begin{aligned}
N_{~t}^t &= N_{~r}^r = 0 \, , \\
N_{~t}^r &= \frac{1}{192\pi}\Big(f^{\prime}h^{\prime} +f^{\prime\prime}h\Big) \, , \\
N_{~r}^t &= \frac{-1}{192\pi h^2}\Big(h^{\prime 2} -h^{\prime\prime}h\Big) \, .
\end{aligned}\ee

Taking into account the time independence of $T_{~\nu}^{\mu}$, we can integrate Eq. (14),
up to a trace $T_{~\alpha}^{\alpha}(r)$, to get
\be\begin{aligned}
T_{~t}^t &= -\frac{K +Q}{f} -\frac{B(r)}{f} -\frac{I(r)}{2f} +T_{~\alpha}^{\alpha}(r) \, , \\
T_{~r}^r &= \frac{K +Q}{f} +\frac{B(r)}{f} +\frac{I(r)}{2f} \, , \\
T_{~t}^r &= -\sqrt{h/f}K +C(r) = -fhT_{~r}^t \, ,
\end{aligned}\ee
where $C(r) = \sqrt{h/f}\int_{r_H}^r \sqrt{f(x)/h(x)}A_t(x)dx$, $B(r) = \int_{r_H}^r A_r(x)f(x)dx$,
$I(r) = \int_{r_H}^r T_{~\alpha}^{\alpha}(x)f^{\prime}(x)dx$, $K$ and $Q$ are two integration constants.
For the line element (11), $B(r)$ should be zero because $A_r = 0$ in the near--horizon region. In the
limit $r \to r_H$, we have $C(r) \to 0$, and $I(r)/f\big|_{r_H} \to T_{~\alpha}^{\alpha}(r_H)$.

Under the infinitesimal general coordinate transformations, the effective action varies as
\bea
&&\hspace*{-0.85cm}
-\delta_{\lambda}W = \int{d^2x\sqrt{-g}\lambda^{\nu}\nabla_{\mu}T_{~\nu}^{\mu}} \nn \\
&&~ = \int dtdr\lambda^{\nu}\Big\{\p_{\mu}\big[N_{~\nu}^{\mu}H(r)\big]
 +\big[\sqrt{f/h}T_{(o)\nu}^{\mu} \nn \\
&&\quad~  -\sqrt{f/h}T_{(H)\nu}^{\mu} +N_{~\nu}^{\mu}\big]\p_{\mu}\Theta(r) \Big\} \nn \\
&&~ = \int dtdr \Big\{\lambda^t \Big(\p_r\big[N_{~t}^rH(r)\big] +\Big[ N_{~t}^r\nn \\
&&\quad~ +\sqrt{f/h}\big(T_{(o)t}^r -T_{(H)t}^r\big) \Big]\delta(r -r_H)\Big) \nn \\
&&\quad~ +\lambda^r\sqrt{f/h}\big(T_{(o)r}^r -T_{(H)r}^r\big)\delta(r -r_H)\Big\} \, ,
\eea
where $\Theta(r) = \Theta(r -r_H -\varepsilon)$ is a scalar step function, $H(r) = 1 -\Theta(r)$
is a scalar top hat function, and we have written the total energy--momentum tensor $T_{~\nu}^{\mu}$
outside the horizon as
\be
T_{~\nu}^{\mu} = T_{(o)\nu}^{\mu}\Theta(r) +T_{(H)\nu}^{\mu}H(r) \, ,
\ee
in which $T_{(o)\nu}^{\mu}$ is covariantly conserved and $T_{(H)\nu}^{\mu}$ obeys the anomalous
Eq. (14). To derive the last expression of Eq. (17), we have taken the $\varepsilon \to 0$ limit.

In order to keep the diffeomorphism invariance, the variation of the effective action should vanish.
The first term $\p_r\big[N_{~t}^rH(r)\big]$ in Eq. (17) can be cancelled by the quantum effects of
the ingoing modes. Setting $\delta_{\lambda}W = 0$, we get the following constrains
\be\begin{aligned}
&\Big[\sqrt{f/h}\big(T_{(o)t}^r -T_{(H)t}^r\big) +N_{~t}^r\Big]\Big|_{r_H} = 0 \, ,\\
&\big(T_{(o)r}^r -T_{(H)r}^r\big)\big|_{r_H} = 0 \, ,
\end{aligned}\ee
i.e.,
\be K_{o} = K_H +\Phi \, , \qquad Q_{o} = Q_H -\Phi \, ,
\ee
where
\be
\Phi = N_{~t}^r(r_H) = \frac{f^{\prime}(r_H)h^{\prime}(r_H)}{192\pi}
= \frac{\kappa^2}{48\pi} \, .
\ee

In order to fix the four constants, we impose an additional regularity condition that requires the
covariant energy--momentum tensor to vanish at the horizon. In the background of space-time (11),
the covariant energy--momentum tensor is related to the consistent one by$^{[8]}$
\be
\sqrt{f/h}\widetilde{T}_{~t}^{r} = \sqrt{f/h}T_{~t}^{r}+\frac{h}{192\pi f}
\Big[ff^{\prime\prime} -2(f^{\prime})^{2}\Big] \, .
\ee
The condition $\widetilde{T}_{~t}^{r}(r_H) = 0$ yields $K_H = -2 \Phi$, which leads to $\sqrt{f/h}
T_{(o)t}^r = -K_{o} = \Phi$. So, $\Phi$ is the flux of Hawking radiation. A $(1+1)$--dimensional
black body radiation has a flux of the form $\Phi = \frac{\pi}{12}T^2$, accurately giving the
Hawking temperature $T = \kappa/(2\pi)$.

Applying the above analysis to the metric (7), we can obtain the correct Hawking temperature
$T = 1/(8\pi\widetilde{m})$; whereas to the line element (1), we will get a different result
$T = (1 -\eta^2)^2/(8\pi m)$. So it is unadvisable to apply directly the RW's method to the
space--time (1). However, we can do the same analysis in another different way. By re-scaling
$t \to \sqrt{1 -\eta^2}~t$, we rewrite the metric (1) as
\be\begin{aligned}
ds^2 &= -f(r)dt^2 +h(r)^{-1}dr^2 +r^2d\Omega_2^2 \, , \\
f(r) &= 1 -\frac{2m}{(1 -\eta^2)r} \, , \quad h(r) = 1 -\eta^2 -\frac{2m}{r} \, ,
\end{aligned}\ee
and immediately derive the correct Hawking temperature $T = (1 -\eta^2)^{3/2}/(8\pi m)$.

In summary, we have applied the RW's method to derive the Hawking flux from a Schwarzschild-type
black hole with the global monopole by requiring the cancellation of gravitational anomalies at
the horizon. The flux has a form precisely equivalent to black body radiation with the Hawking
temperature. To obtain the consistent expression of the Hawking temperature, it is not suitable to
use the metric (1), otherwise one must divide the flux $\Phi$ by a factor $1 -\eta^2$. Our analysis
presented here can be directly applied to the case of a Schwarzschild-anti-de Sitter black hole with
a global monopole where $f(r) = 1 -\eta^2 -2m/r +r^2/l^2$.


\emu
\bc \rule{8cm}{0.1mm} \ec
\bmu
\vs{3mm}

\emu
\end{document}